\documentclass{ws-ijmpa}

\usepackage[super]{cite}
\usepackage{xcolor}
\usepackage[verbose,hypertexnames=false]{hyperref}
\hypersetup{colorlinks=false,allbordercolors=blue,pdfborderstyle={/S/U/W 1}}

\usepackage{pdflscape}
\usepackage{amsmath}
\usepackage{amsfonts}
\usepackage{amssymb}
\usepackage{graphicx}
\usepackage[titletoc]{appendix}
\usepackage{color}
\usepackage{hyperref}
\usepackage{cleveref}
\usepackage[rightcaption]{sidecap}
\usepackage{comment}
\usepackage{soul}
\usepackage{cancel}

\usepackage{dcolumn}

\usepackage{longtable}
\usepackage{floatrow}
\usepackage{array}
\usepackage{ctable}
\usepackage{multirow}
\usepackage{siunitx}
\usepackage{tabularx}
\usepackage{booktabs}
\usepackage{supertabular}

\graphicspath{{Graphics/}}

\def\be{\begin{equation}}
\def\ee{\end{equation}}
\def\bea{\begin{eqnarray}}
\def\eea{\end{eqnarray}}

\definecolor{vividviolet}{rgb}{0.62, 0.0, 1.0}
\definecolor{amaranth}{rgb}{0.9, 0.17, 0.31}
\definecolor{palatinateblue}{rgb}{0.15, 0.23, 0.89}
\definecolor{brightpink}{rgb}{1.0, 0.0, 0.5}
\definecolor{cornflowerblue}{rgb}{0.39, 0.58, 0.93}
\definecolor{deepcarminepink}{rgb}{0.94, 0.19, 0.22}
\definecolor{radicalred}{rgb}{1.0, 0.21, 0.37}

\hypersetup{ linktoc=all,
    colorlinks, linkcolor={palatinateblue},
    citecolor={brightpink}, urlcolor={amaranth}
}

\begin{document}

\markboth{Y. Carloni}{Alleviating the Hubble tension with the $\Lambda_{\omega_s}$CDM model}

%
\catchline{}{}{}{}{}
%

\title{Alleviating the Hubble tension with the $\Lambda_{\omega_s}$CDM model}

\author{Youri Carloni}

\address{School of Science and Technology, University of Camerino, Via Madonna delle Carceri, Camerino, 62032, Italy,
INAF - Osservatorio Astronomico di Brera, Milano, Italy, 
and Istituto Nazionale di Fisica Nucleare (INFN), Sezione di Perugia, Perugia, 06123, Italy.
\\E-mail: youri.carloni@unicam.it}

\maketitle

\begin{abstract}
We formulate a novel extension of the $\Lambda$CDM model, named $\Lambda_{\omega_s}$CDM, in which we consider an additional term at early times in order to alleviate the Hubble tension. This additional component, referred to as \emph{matter with pressure}, indicates a barotropic fluid that is subdominant to dust and radiation as the Universe expands, thereby recovering the $\Lambda$CDM paradigm at late times. We constrain the $\Lambda_{\omega_s}$CDM cosmology by performing a Markov Chain Monte Carlo analysis with Planck 2018 CMB, DESI DR2, and Pantheon+\texttt{SH0ES} data. The results suggest that the barotropic factor and the normalized density of the new fluid are given, respectively, by $\omega_s=0.294_{-0.004(0.023)}^{+0.014(0.015)}$ and $10^{5}\Omega_s=1.62_{-0.56(0.91)}^{+0.36(1.02)}$. With these two additional parameters, the Hubble constant is increased to
$H_0 = 71.51^{+0.72(1.43)}_{-0.74(1.46)}$~km/s/Mpc, alleviating \emph{de facto} the Hubble tension.
\end{abstract}



\section{Introduction}

The $\Lambda$CDM model is the simplest way to describe the Universe and is confirmed by several observations \cite{l1,l2,l3,l4}.

However, recently DESI results suggest that the dynamical dark energy, under the form of the $\omega_0\omega_a$CDM parameterization, is favored over the cosmological constant \cite{desi2024, desi2025,desi1,desi2,desi3,desi4,desi5,desi6}. In addition, the concordance paradigm suffers from several cosmological tensions, such as the $S_8$ and Hubble tensions \cite{S8tension,H0tension}.

The Hubble tension arises from the discrepancy between indirect and direct measurements of the Hubble constant $H_0$. In particular, assuming the $\Lambda$CDM model, we have $H_0=(67.36\pm0.54)$~km/s/Mpc from CMB data \cite{l4}, while considering local measurements from \texttt{SH0ES} collaboration $H_0=(73.04\pm1.04)$~km/s/Mpc \cite{H0Riess}. This indicates a $4.1\sigma$ tension between the two methodologies, providing a possible hint of new physics. Consequently, several models have been proposed to alleviate this tension.

The approaches are generally classified into two categories: late-time and early-time solutions. Late-time solutions aim to increase the Hubble constant by reducing the energy content of the Universe at late times \cite{latet1, latet2}. In contrast, early-time solutions consist of reducing the comoving sound horizon by increasing the Hubble parameter at early times \cite{early1, early2}. 

Although both approaches remain viable, recent analyses favor early-time scenarios. Indeed, some studies suggest that the Hubble tension may originate from a calibration discrepancy \cite{nolate1, nolate2}, and therefore resolving the tension requires modifying the comoving sound horizon.

In this context, we introduce an extension of the $\Lambda$CDM model, denoted as $\Lambda_{\omega_s}$CDM, aimed at alleviating the Hubble tension \cite{lwscdm,lwscdm2}. Specifically, we consider an additional barotropic fluid that contributes only at early times, leaving the late-time $\Lambda$CDM evolution unchanged. We test this scenario through a Markov Chain Monte Carlo analysis combining Planck 2018 CMB, DESI DR2, and Pantheon+\texttt{SH0ES} data. The results show that our model reduces the $H_0$ tension from $4.1 \sigma$ to $2.4 \sigma$, and the Bayesian evidence reveals a strong statistical preference for $\Lambda_{\omega_s}$CDM cosmology over the concordance paradigm.

\section{$\Lambda_{\omega_s}$CDM cosmology}

\subsection{The early Universe with the new fluid}

We consider an early-time barotropic component, dubbed \emph{matter with pressure}. Thus, before the recombination, the Universe is composed of:
\begin{itemize}
    \item[-] photons, with equation of state $\omega_\gamma=1/3$;
    \item[-] baryons, with equation of state $\omega_b \simeq 0$;
    \item[-] matter with pressure, with a constant equation of state $\omega_{s}$.  
\end{itemize}
The additional component modifies the comoving sound horizon $r_s$, defined as
\begin{equation}
r_s(z_\star) = \int_{z_\star}^{\infty} \frac{c_s(z)}{H(z)}dz,
\end{equation}
with $z_\star$ representing the redshift of recombination. The fluid affects both the sound speed and the Hubble expansion rate prior to recombination. 

In particular, the sound speed is given by
\begin{equation}
c_s = c\sqrt{\frac{1 + 3\omega_s W}{3\left(1 + R + W\right)}},
\end{equation}
where $R\equiv 3\rho_b/(4\rho_\gamma)$, and $W \equiv 3(1+\omega_s)\rho_s/(4\rho_\gamma)$, with $\rho_b$, $\rho_s$, and $\rho_\gamma$ denoting the energy densities of baryons, matter with pressure and photons.

The Hubble parameter instead takes the form
\begin{equation}
H(z) = H_0 \sqrt{\Omega_{\Lambda}+\Omega_m(1+z)^3+\Omega_r(1+z)^4+\Omega_s(1+z)^{3(1+\omega_s)}} ,
\label{eq:hz}
\end{equation}
where $\Lambda,m,r,s$, label the cosmological constant, dust, radiation and matter with pressure, respectively.

Then, we assume that the new fluid contributes positively to the energy–momentum tensor, in analogy with standard matter components and in contrast to dark energy. This requirement implies the Zeldovich limit, $\omega_s > 0$. 

In addition, the fluid should decay faster than cold dark matter and baryons as the Universe expands. As a consequence, matter with pressure increases the total energy content at early times, raising the inferred value of $H_0$, and it becomes negligible at late times, recovering the $\Lambda$CDM paradigm.

\subsection{The combined analysis}

We tested our new scenario by using a modified version of the \texttt{CLASS} code \footnote{\url{https://github.com/YouriCarloni/LwsCLASS}}. The model parameters are then constrained through a  Markov Chain Monte Carlo analysis performed with \texttt{MontePython}.

Our analysis employs the Pantheon+ Type Ia supernovae sample with \texttt{SH0ES} 
Cepheid host distance anchors providing the $H_0$ calibration, the DESI DR2 baryon 
acoustic oscillation measurements, and Planck 2018 CMB data, specifically the 
high-$\ell$ \texttt{TTTEEE}, low-$\ell$ \texttt{TTEE}, and lensing likelihoods.

We assume convergence of the chains when the Gelman–Rubin statistic satisfies $R - 1 < 0.02$.

\begin{figure}[htb!]
\centering
\includegraphics[width=0.95\hsize,clip]{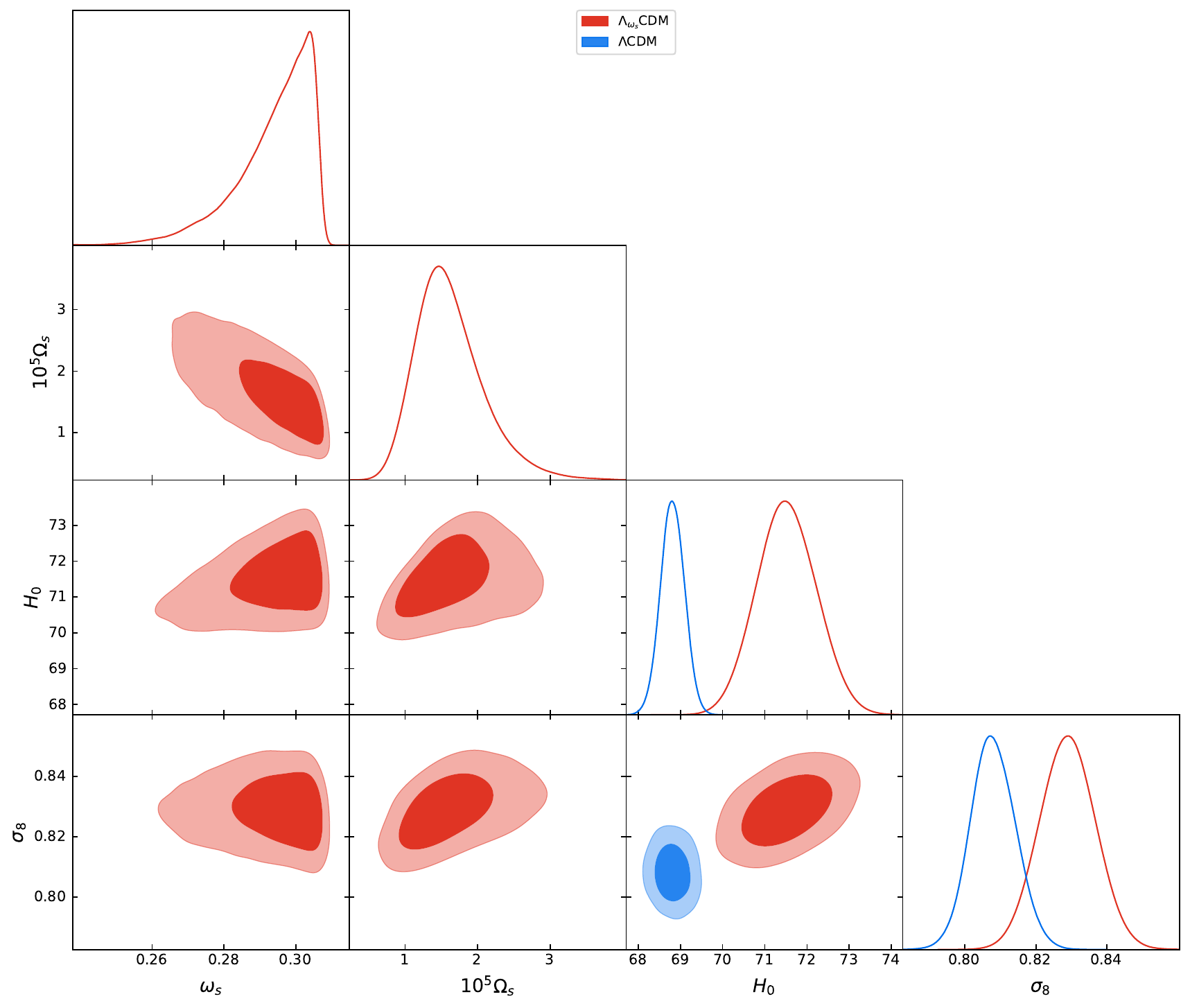}
\caption{Posteriors for the $\Lambda_{\omega_s}$CDM and $\Lambda$CDM  models with respect to the Planck 2018 CMB, DESI DR2, and Pantheon+\texttt{SH0ES} data.}
\label{fig:tri}
\end{figure}

The results obtained from the combined analysis are given in Tab.~\ref{tab:best}, and the contour plots are displayed in Fig.~\ref{fig:tri}.

\begin{table}[htb!]
\tbl{Mean values and $1\sigma$ ($2\sigma$) confidence intervals for the cosmological parameters of the $\Lambda_{\omega_s}$CDM and $\Lambda$CDM models obtained from the Markov Chain Monte Carlo analysis. We also report the $\chi^2$ and $\Delta\chi^2$, computed relative to the $\Lambda$CDM model, together with the Bayesian evidence $\Delta \log B$.}
{\begin{tabular}{lcc}
\hline\hline
    Parameter &~~$\Lambda_{\omega_{s}}$CDM~~&~~~$\Lambda$CDM~~~\\ 
    \hline
    $100~\omega_b$ & $2.271^{+0.015(0.029)}_{-0.014(0.029)}$ &  $2.265^{+0.013(0.026)}_{-0.013(0.026)}$ \\
    $\omega_{\rm cdm}$&$0.1278^{+0.0027(0.0055)}_{-0.0028(0.0054)}$ & $0.1171^{+0.0006(0.0013)}_{-0.0006(0.0013)}$ \\
     $H_0$~(km/s/Mpc) & $71.51^{+0.72(1.43)}_{-0.74(1.46)}$ &$68.88^{+0.29(0.58)}_{-0.29(0.58)}$    \\
    $10^{9}A_s$ &$2.122^{+0.031(0.069)}_{-0.036(0.067)}$ & $2.123^{+0.032(0.067)}_{-0.034(0.066)}$\\
    $n_s$& $0.9814_{-0.0048(0.0086)}^{+0.0040(0.0090)}$  &$0.9728_{-0.0039(0.0069)}^{+0.0034(0.0069)}$ \\
    $\tau_{\rm reio}$ &  $0.060^{+0.008(0.016)}_{-0.007(0.015)}$ &$0.064^{+0.007(0.016)}_{-0.008(0.015)}$ \\
    $\omega_s$ &$0.294_{-0.004(0.023)}^{+0.014(0.015)}$  &$-$ \\
    $10^{5}\Omega_s$  & $1.62_{-0.56(0.91)}^{+0.36(1.02)}$ & $-$ \\
    $r_s(z_{\star})$~(Mpc) &$139.9^{+1.2(2.4)}_{-1.2(2.4)}$ & $145.1^{+0.2(0.4)}_{-0.2(0.3)}$\\
    $\sigma_8$ & $0.829^{+0.008(0.016)}_{-0.008(0.016)}$ &$0.808^{+0.006(0.013)}_{-0.007(0.013)}$ \\
     $\Omega_m$ & $0.296^{+0.004(0.007)}_{-0.004(0.007)}$&$0.296^{+0.004(0.007)}_{-0.004(0.007)}$ \\
    $S_8$ & $0.823_{-0.010(0.019)}^{+0.009(0.019)}$&$0.803^{+0.008(0.016)}_{-0.008(0.016)}$  \\
     $100~\theta_s$ & $1.0433^{+0.0004(0.0008)}_{-0.0004(0.0008)}$ & $1.0422^{+0.0003(0.0006)}_{-0.0003(0.0006)}$ \\
    \hline 
    $\chi^{2}(\Delta\chi^{2})$ & $4093.06(-22.42)$ & $4115.48(0)$ \\
     $\Delta \log B$ & $+6.2$ & $0$ \\
     \hline
\end{tabular}
\label{tab:best}}
\end{table}

We then assess whether our model provides a better description than the concordance paradigm by computing the Bayesian evidence using the public code \texttt{MCEvidence}\footnote{\url{https://github.com/yabebalFantaye/MCEvidence}}.

We find that $\Delta \log B = \log B_{\rm s}-\log B_\Lambda=+6.2$, showing a decisive statistical preference in favor of $\Lambda_{\omega_s}$CDM according to the modified Jeffreys’ scale.

We therefore conclude that introducing matter with pressure at early times can alleviate the $H_0$ tension. Specifically, the $\Lambda_{\omega_s}$CDM model reduces the tension from $4.1 \sigma$ to $2.4 \sigma$. Furthermore, the mean values of the additional cosmological parameters suggest that the fluid driving the increase in $H_0$ behaves similarly to radiation. Indeed, from Tab.~\ref{tab:best}, we can derive that
\begin{equation}
\omega_s=\frac{1}{3}-\epsilon,\qquad \Omega_s(z)=\Omega_s(1+z)^{4-3\epsilon},
\end{equation}
with $\epsilon=0.040^{+0.003}_{-0.009}$. 

From a physical point of view, our new fluid can be interpreted as a thermalized scalar field, whose dynamics may be associated with quasi-quintessence, or as a vector field, with the Proca field providing a natural and plausible realization.

\begin{figure}[htb!]
\centering
\includegraphics[width=0.75\hsize,clip]{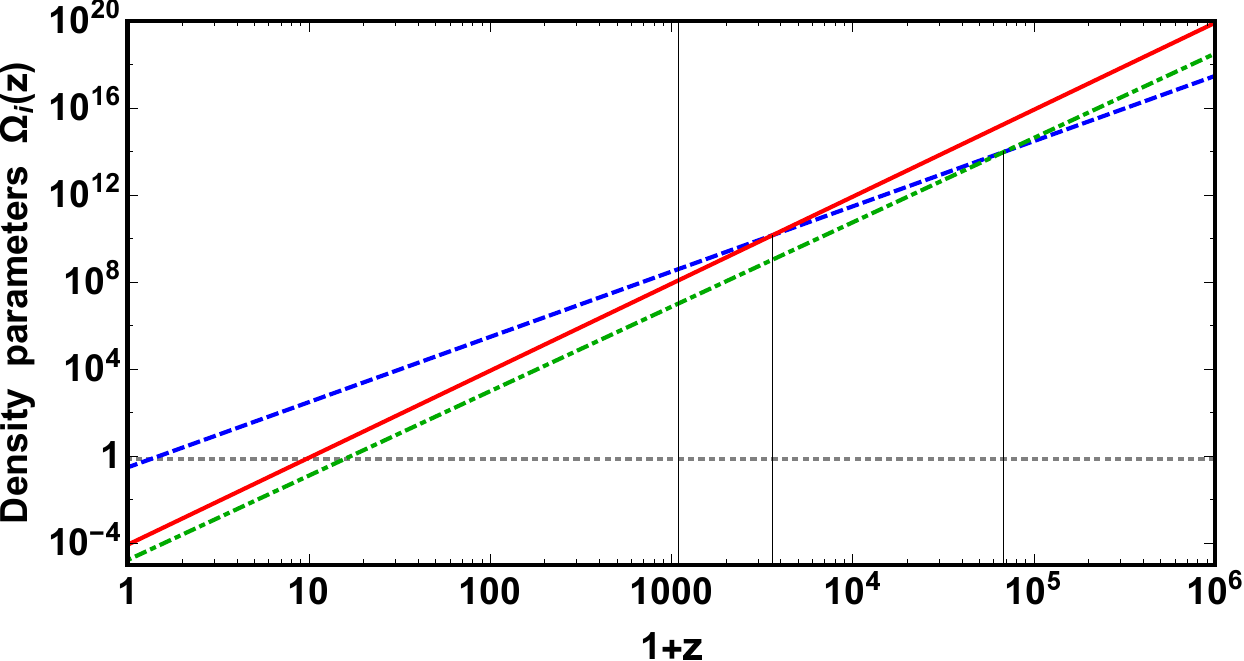}
\caption{Evolution of the energy densities in the $\Lambda_{\omega_s}$CDM cosmology: $\Lambda$ (dotted gray), dust (dashed blue), radiation (solid red), and matter with pressure (dot--dashed green). Vertical black lines indicate, from left to right, recombination, dust--radiation equality, and dust--matter with pressure equality.}
\label{fig:eq}
\end{figure}

Finally, using the barotropic factor and the normalized density obtained from the analysis, we derive the density evolution of the new component and compare it with that of dust and radiation, as shown in Fig.~\ref{fig:eq}. 

Here, we find that matter with pressure is subdominant with respect to dust and radiation at late times, recovering the $\Lambda$CDM cosmology.

\section{Conclusions and outlook}

In this work, we proposed a novel early-time solution to the Hubble tension, named $\Lambda_{\omega_s}$CDM cosmology. The model introduces an additional barotropic fluid at early times that reduces the comoving sound horizon and, as a consequence, increases the Hubble constant. Specifically, we extend the $\Lambda$CDM paradigm by adding two more parameters, i.e., the barotropic factor $\omega_s$ and the normalized density $\Omega_s$. Then, we perform a Markov Chain Monte Carlo analysis by combining the Planck 2018 CMB, DESI DR2 and Pantheon+\texttt{SH0ES} data. The results suggest that the characteristics of our new fluid are similar to those of radiation, leading to $\omega_s=0.294_{-0.004(0.023)}^{+0.014(0.015)}$, and $10^{5}\Omega_s=1.62_{-0.56(0.91)}^{+0.36(1.02)}$. With these new parameters, the Hubble constant is raised to $H_0 = 71.51^{+0.72(1.43)}_{-0.74(1.46)}$~km/s/Mpc, alleviating the tension from $4.1 \sigma$ to $2.4 \sigma$ confidence level. Thus, we employ a new model in which the additional fluid, denoted as matter with pressure, is significant at early times and becomes negligible as the Universe expands.

Finally, we compare the $\Lambda_{\omega_s}$CDM cosmology and the concordance paradigm from a statistical point of view through the Bayesian evidence, and we find a decisive preference for our model. 

Future work will focus on a more detailed investigation of the physical properties of matter with pressure, and a comparison with various early dark energy models will also be carried out to assess the competitiveness of our new scenario.

\end{document}